\documentclass[aps,twocolumn,prd,showpacs,nofootinbib,preprintnumbers]{revtex4}
\usepackage{amsmath}
\usepackage{graphicx}
\usepackage{subfigure}
\usepackage{dcolumn}
\usepackage{bm}
\usepackage{amssymb}
\usepackage{latexsym}
\usepackage{psfrag}

\newcommand{\setC}{\mathbb{C}}

\newcommand{\setR}{\mathbb{R}}

\newcommand{\ie}{\emph{i.e.~}}
\newcommand{\GReCO}{${\cal G}\setR\varepsilon\setC{\cal O}$}

\newcommand{\df}{\delta\varphi}
\newcommand{\Df}{\Delta\varphi}

\newcommand{\mean}[1]{\left\langle #1 \right\rangle}
\newcommand{\mpl}{m_{_\mathrm{Pl}}}
\newcommand{\MPl}{M_{_\mathrm{Pl}}}
\newcommand{\ini}{\mathrm{in}}\newcommand{\cl}{\mathrm{cl}}

\newcommand{\dd}{\mathrm{d}}

\renewcommand{\t}{\tau}
\newcommand{\e}{\mathrm{e}}

\begin{document}

\title{On the Reliability of the Langevin Pertubative Solution in
  Stochastic Inflation}

\author{J\'er\^ome Martin} \email{jmartin@iap.fr}
\affiliation{Institut d'Astrophysique de Paris, \GReCO, UMR 7095-CNRS,
Universit\'e Pierre et Marie Curie, 98bis boulevard Arago, 75014
Paris, France}

\author{Marcello Musso} \email{musso@physics.utexas.edu}
\affiliation{University of Texas at Austin, Department of
Physics - Theory group, 1 University Station C1608, Austin TX
78712-0269 USA}

\preprint{UTTG-15-05}

\date{\today}

\begin{abstract}
A method to estimate the reliability of a perturbative expansion of
the stochastic inflationary Langevin equation is presented and
discussed. The method is applied to various inflationary scenarios, as
large field, small field and running mass models. It is demonstrated
that the perturbative approach is more reliable than could be naively
suspected and, in general, only breaks down at the very end of
inflation.
\end{abstract}

\maketitle

%%%%%%%%%%%%%%%%%%%%%%%%%%%%%%%%%%%%%%%%%%%%%%%%%%%%%%%%%%%%%%%%%%%
%%%%%%                    Introduction                       %%%%%%
%%%%%%%%%%%%%%%%%%%%%%%%%%%%%%%%%%%%%%%%%%%%%%%%%%%%%%%%%%%%%%%%%%%

\section{Introduction}

The stochastic approach to inflation~\cite{early,stocha,NNS,NS,N} is
an efficient method to study how quantum effects can influence the
dynamics of the scalar field driving the acceleration of the early
Universe. This formalism is based on a Langevin equation which
describes the evolution of a spatially averaged field (typically over
a Hubble patch), the so-called ``coarse-grained'' field. Solving this
equation, especially when the backreaction of this coarse-grained
field on the background geometry is taken into account, is notoriously
known as a difficult task and various methods have been proposed in
the literature, see for instance Refs.~\cite{Hodges,scaling,YVM,MM1}.

\par

Recently, a method based on a perturbative expansion in the noise was
presented~\cite{MM2}. At second order in the noise, this method is
powerful enough to ensure the calculation of the probability density
function of the coarse-grained field for arbitrary potentials.  It was
demonstrated that, in order to obtain explicit analytical expressions,
the calculation of only one quadrature is necessary.  If, in addition,
the volume effects are also determined, then only one more quadrature
is required. It turns out that these quadratures are feasible for a
large class of inflationary models, for instance in the cases of the
chaotic~\cite{chaotic}, new~\cite{new}, hybrid~\cite{hybrid} and
running mass~\cite{running} scenarios. The stochastic effects in these
models were studied in Ref.~\cite{MM2}, where the evolution of the
corresponding distributions was discussed in detail.

\par

An important question concerns the domain of validity of the
perturbative expansion used in order to obtain the above results. The
aim of this article is to develop a method to treat this question and
to estimate when the perturbative expansion gives reliable results. It
is worth noticing that so far (and this is also valid for the other
approaches used in the literature to solve the Langevin equation) this
issue has never been addressed elsewhere. In general, the approximate
expression for the probability function is derived without worrying
about its accuracy. It will be shown that the method of
Ref.~\cite{MM2} gives, most of the time, a better approximation than
it could naively be guessed on general grounds and only breaks down at
the very end of inflation.

\par

Our article is organized as follows. In the next section,
Sec.~\ref{Solving the Langevin Equation}, we briefly recall the main
results and equations obtained in Ref.~\cite{MM2}. Then, in
Sec.~\ref{Reliability of the Expansion}, we present our method to
study the accuracy of the perturbative expansion and apply it to the
inflationary models discussed in Ref.~\cite{MM2}, to wit, chaotic,
new, hybrid and running mass scenarios. Finally, in
Sec.~\ref{Conclusions}, we present our conclusions.

\section{Solving the Langevin Equation} 
\label{Solving the Langevin Equation}

In stochastic inflation, one is interested in the behavior of a
coarse-grained field $\varphi $ obtained after taking the spatial
average of the original inflaton field over a volume the size of which
is of the order of a Hubble patch. The coarse-grained field obeys a
Langevin equation that can be written as
\begin{equation}
\label{Langevin}
  \dot\varphi +\frac{1}{3H}\frac{{\rm d}V}{{\rm
  d}\varphi}= \frac{H^{3/2}}{2\pi}\xi(t)\, ,
\end{equation}
where $V$ is the inflaton potential and $\xi $ a white noise defined
such that its correlation function simply reads $
\left\langle\xi(t)\xi(t')\right\rangle=\delta(t-t')$, $\delta(z)$
being the Dirac distribution. The backreaction can be seen in the fact
that the Hubble parameter $H$ in Eq.~(\ref{Langevin}) depends on the
coarse-grained field $\varphi $ via the slow-roll Friedmann equation
$H^2\sim \kappa V(\varphi )/3$, where $\kappa =8\pi /\mpl^2$.

\par

The method proposed in Ref.~\cite{MM2} (see also Ref.~\cite{GLMM} for
earlier attempts) consists in expanding the coarse-grained field in
powers of the noise according to
\begin{equation}
  \varphi(t) = \varphi_\cl(t)+\df_1(t)+\df_2(t)+\cdots \, ,
\end{equation}
where $\varphi _\cl$ is the classical solution, \ie the one obtained
when the noise is ``switched off'' in the Langevin equation. The
quantities $\df_1$ and $\df_2$ are respectively first and second order
in the noise. They obey the equations
\begin{equation}
\label{eqdefi1}
  \frac{{\rm d}\df_1}{{\rm d}t} + \frac{2}{\kappa}H''(\varphi_{\rm
  cl}) \df_1 = \frac{H^{3/2}(\varphi _{\rm cl})}{2\pi}\xi(t)\, ,
\end{equation}
and
\begin{align}
\label{eqdefi2}
  & \frac{{\rm d}\df_2}{{\rm d}t} + \frac{2}{\kappa}H''(\varphi
  _{\rm cl}) \df_2 = -\frac{H'''(\varphi _{\rm cl})}{\kappa}\df_1^2
  \nonumber \\ & + \frac{3}{4\pi}H^{1/2}(\varphi _{\rm
  cl})H'(\varphi _{\rm cl})\df_1\xi(t)\, ,
\end{align}
where a prime denotes a derivative with respect to the field. These
equations can easily be solved since they are (by definition)
linear. The expansion could of course be pushed to higher orders if
necessary.

\par

In Ref.~\cite{MM2}, it was demonstrated that this formalism allows us
to calculate the probability density function $P_{\rm c}$ of the
coarse-grained field in a single Hubble patch. At second order in the
noise, it is given by
\begin{equation}
\label{pc}
  P_{\rm c}(\varphi,t) = \frac{1}{\sqrt{2\pi\!\mean{\df_1^2}}}
  \exp\!\left[
  -\frac{(\varphi-\varphi_\cl-\mean{\df_2})^2}{2\mean{\df_1^2}}\right]\,
  , 
\end{equation}
where the variance $\mean{\df_1^2}$ and the mean $\mean{\df_2}$
appearing in the above expression can respectively be written as
\begin{align}
\label{slowvar}
  \mean{\df_1^2} &= \frac{\kappa}{2}\left(\frac{H'}{2\pi}\right)^2
  \!\!\int^{\varphi_\ini}_{\varphi _{\rm cl}}\!\!\dd\psi
  \left(\frac{H}{H'}\right)^3\, , \\
\label{slowmean}
  \mean{\df_2} &= \frac{H''}{2H'}\mean{\df_1^2}
  +\frac{H'}{4\pi\mpl^2}\Bigg[\frac{H_\ini^3}{(H_\ini')^2}
  -\frac{H^3}{(H')^2}\Bigg]\, .
\end{align}

\par

This formalism also permits the calculation of volume effects. If,
instead of considering the distribution of the field in a single
domain, we want to have access to its spatial distribution, one must
weigh the single domain distribution by the volume of each Hubble
patch. This leads to the definition
\begin{equation}
\label{volume}
  P_{\rm v}(\varphi,t) =
  \frac{\mean{\delta(\varphi-\varphi[\xi])\,\e^{3\!\int\dd\t
  H(\varphi[\xi])}}} {\mean{\e^{3\!\int\dd\t H(\varphi[\xi])}}}\, .
\end{equation}
Then, it was shown in Ref.~\cite{MM2} that, at second order in the
noise, $P_{\rm v}$ takes the form
\begin{equation}
\label{pv}
  P_{\rm v}(\varphi,t) = \frac{1}{\sqrt{2\pi\!\mean{\df_1^2}}}
  \exp\!\left[ -\frac{\left(\varphi-\mean{\varphi}-\!3I^{\rm
  T}\!\!J\right)^2}{ 2\mean{\df_1^2}}\right]\, ,
\end{equation}
where $\mean{\df_1^2}$ and $\mean{\varphi}=\varphi_\cl+\mean{\df_2}$
are still given by Eqs.~(\ref{slowvar}) and (\ref{slowmean}). The term
$I^{\rm T}\!J$ describing the correction to the mean value due to
volume effects can be written as
\begin{equation}
\label{voleffects}
  3I^{\rm T}\!J= \frac{12H'}{\mpl^4}
  \!\int_{\varphi_\cl(t)}^{\varphi_\ini}\!\!\dd\psi\frac{H^4}{(H')^3}
  -12\pi\frac{H}{H'}\frac{\mean{\df_1^2(t)}}{\mpl^2}\, .
\end{equation}
Therefore, as already mentioned, estimating the volume effects merely
requires the calculation of one additional quadrature.

\par

In Ref.~\cite{MM2}, the results briefly described above have been
applied to various concrete inflationary models. In particular, the
potential 
\begin{equation}
\label{potential}
  V(\varphi) = M^4\!\left[a + b
  \left(\frac{\varphi}{\mu}\right)^{\!\!n}\right]\, ,
\end{equation}
where $a=0,1$ and $b=\pm1$, has been considered. The case $a=0$, $b=1$
corresponds to large field (LF) models (or ``chaotic
inflation'')~\cite{chaotic}, $a=1$, $b=-1$ to small field (SF) models
(as ``new inflation'')~\cite{new} and $a=1$, $b=1$ to hybrid
inflation~\cite{hybrid}. The scale $M$ is fixed by the Wilkinson
Microwave Anisotropy Probe (WMAP) normalization. The case of running
mass (RM) inflation~\cite{running}, namely
\begin{equation}
\label{potentialrunning}
  V(\varphi) = M^4\left[1-\frac{c}{2}\left(-\frac{1}{2} +\ln
\frac{\varphi }{\varphi _0}\right)\frac{\varphi ^2}{M_{_{\rm
Pl}}^2}\right],
\end{equation}
was also treated. In the expression of the potential, $M_{_{\rm
Pl}}\equiv \mpl/\sqrt{8\pi }$ and the quantities $c$ (which can be
positive or negative) and $\varphi _0$ are free parameters. Running
mass inflation can be realized in four classical versions and
stochastic effects have been studied in Ref.~\cite{MM2} for the first
($c>0$, $\varphi _\cl<\varphi_0$) and the second ($c>0$, $\varphi _\cl
>\varphi _0$) scenarios (RM1 and RM2) .

\par

For the models described above, the behavior of $P_{\rm c}$ and
$P_{\rm v}$ have been investigated in details in Ref.~\cite{MM2}, see
in particular Figs.~2 and~3. As mentioned in the introduction, the
issue that we now address is the reliability of the method of
approximation used in order to establish these results.

\section{Reliability of the Expansion}
\label{Reliability of the Expansion}

An important question is the determination of the interval in which
the perturbed solution of the Langevin equation that we have obtained,
$\varphi_\cl+\df_1+\df_2$, remains a good approximation of the exact
one.  Indeed, initially, the perturbed solution is ``by definition'' a
good approximation since we have $\df_1(\varphi_\ini)=\df_2(\varphi
_\ini)=0$. Then, as the field evolves from $\varphi_\ini$, we expect
$\df_1 $ and $\df_2$ to grow and the approximation to break down at
some value of $\varphi _\cl\neq \varphi _\ini$. {\it A priori}, the
criterion of validity is simply $\df_2<\df_1<\varphi _\cl$. But things
can be more complicated. For instance, let us assume that the
classical field is initially very small, as is the case for new
inflation. Then, $\df_1/\varphi _\cl $ becomes large very quickly
(because $\varphi _\cl $ is very small), apparently signaling a
breakdown of the approximation. However, it is clear that this could
just be an artifact of the criterion used which, somehow, would be too
naive. To illustrate this last point, let us consider the following
simple example. Suppose that we want to calculate $f(\varphi _\cl
+\Delta \varphi )$, where $f$ is a given function that we do not need
to specify explicitly. Taylor expanding this expression leads to
$f(\varphi _\cl +\Delta \varphi )\sim f(\varphi _\cl)+f'(\varphi
_\cl)\Delta \varphi $ and, in general, this expression gives a good
approximation provided that $\Delta \varphi \ll \varphi
_\cl$. However, if the derivatives of $f$ are very small around
$\varphi _\cl $, then the approximation can be good even if $\Delta
\varphi $ is much larger than could naively be expected. We will see
that, in the case of the perturbative expansion of the Langevin
equation, we are exactly in this situation. The deep reason for that
is the fact that the role of the derivatives of $f$ is now played by
the derivatives of the Hubble parameter. Since these ones are
necessarily small as long as the slow-roll approximation is satisfied,
one can expect the previous phenomenon to happen. Therefore, it is
important not to underestimate the reliability of the perturbative
expansion and to study this issue carefully.

\par

In the following, we address this question from a slightly different
point of view, focusing on the Langevin equation itself rather than on
its exact solution which is, of course, unknown. Our goal is to find a
criterion that controls when the perturbed equation that we are able
to solve is a good approximation of the exact one. This is a simpler
task since we now compare known ``objects''. At this point, one can
even dare an analogy. The situation under consideration is indeed
similar to what is done with the slow-roll approximation, for instance
for the Klein-Gordon equation. In this case, one does not compare the
exact solution (which is, most of the time, unknown as well) to the
slow-roll one. One rather studies how small the term that we neglect
in the exact equation ($\ddot{\varphi}$) is in comparison with the
term that we keep ($H\dot{\varphi }$), \ie we study the magnitude of
the slow-roll parameter $\ddot{\varphi }/(H\dot{\varphi })$. The
spirit of the method that we use below is along the same
line. Finally, before embarking on the discussion of the reliability
of the approximation used here, let us stress again that, so far and
despite its importance, this question has not been given a
satisfactory answer in the literature on the subject.

\par

In order to determine the accuracy of the expansion, we will make use
of the Lagrange remainder theorem \cite{lagr} for the error in a Taylor
expansion. This theorem states that any function
$f(\varphi)$ around some value $\varphi _\cl$ can be written as
\begin{eqnarray}
\label{Lagrange}
  f(\varphi _\cl+\Delta \varphi) &=& \sum_{k=0}^{n-1}
  \frac{f^{(k)}(\varphi _\cl)}{k!}(\Delta \varphi )^k \nonumber \\ & &
  +\frac{f^{(n)}(\varphi_\cl+\theta\Delta \varphi )}{n!}(\Delta
  \varphi )^n \, ,
\end{eqnarray}
for some value of the parameter $\theta$ between $0$ and $1$. Let us
emphasize that this expression is exact and does not assume anything
on $\Delta\varphi$, in particular does not assume $\Delta\varphi\ll
1$.

\par

The next step is to apply this theorem to the Langevin equation,
$\dot{\varphi }+2H'/\kappa =H^{3/2}\xi /(2\pi )$, more precisely to
the function $H'$ and $H^{3/2}$ in the left and right hand sides
respectively. In our case, we take $n=3$ since we have considered the
perturbative expansion of the Langevin equation up to second order in
the noise. This gives
\begin{eqnarray}
\frac{\dd\Df}{\dd t} &+& \frac{2}{\kappa}H''_\cl \,\Df
  +\frac{H'''_\cl}{\kappa}\,\Df^2 +\frac{2L_2}{\kappa }=
  \frac{H^{3/2}_\cl}{2\pi}\,\xi \nonumber \\ &+& \frac{3}{4\pi}H'_\cl
  H^{1/2}_\cl \,\Df\,\xi +\frac{R_2}{2\pi }\,\xi \, ,
\label{eqexact}
\end{eqnarray}
where we have used the classical equation of motion and where,
according to Eq.~(\ref{Lagrange}), we have
\begin{align}
\label{leftcond}
  L_2 &\equiv \frac{H^{(4)}(\varphi_\cl +\theta_{\rm L} \Df)}{6}(\Df)^3 \,,\\
\label{rightcond}
  R_2 &\equiv \frac{\left(H^{3/2}\right)''(\varphi _\cl +\theta_{\rm
  R} \Df)}{2}(\Df)^2\, .
\end{align}
We stress again that, despite its resemblance with
Eqs.~(\ref{eqdefi1}) and~(\ref{eqdefi2}), Eq.~(\ref{eqexact}) is an
exact equation determining $\Delta \varphi $ (hence the exact
stochastic field $\varphi _\cl +\Delta \varphi $), as long as some
values of the two parameters $\theta_{\rm L}$ and $\theta_{\rm R}$ are
suitably chosen between $0$ and $1$. At this stage, this is just a
complicated way to re-write the exact Langevin
equation~(\ref{Langevin}).

\par

The main idea is now to assume that the truncated expansion is
reliable for values of $\Delta \varphi \equiv \df_1 +\df_2$ such that
$L_2$ and $R_2$ are small in comparison with the other terms appearing
in Eq.~(\ref{eqexact}). Indeed, if this is the case, then the
approximated Eqs.~(\ref{eqdefi1}) and (\ref{eqdefi2}) become
indistinguishable from the exact one~(\ref{eqexact}). More precisely,
for each value of $\varphi _\cl $, we have to find the limiting values
$\Df_{\min}(\varphi_\cl)<0$ and $\Df_{\max}(\varphi_\cl)>0$ such that
$L_2$ and $R_2$ are small in comparison with the other terms in the
equation of motion. Then, the validity of the perturbative treatment
will be guaranteed as long as
\begin{equation}
-\vert \Df_{\min}(\varphi _\cl)\vert <\df_1+\df_2<\Df_{\max}(\varphi
  _\cl)\, ,
\end{equation}
or, in other words, as long as we have $\varphi \in \left[\varphi
  _\cl-\left\vert \Df_{\min}(\varphi _\cl)\right\vert , \varphi
  _\cl+\Df_{\max}(\varphi _\cl)\right]$. In practice, since we are
  dealing with stochastic quantities, instead of $\df_1 +\df_2$, we
  will apply our criterion to the quantity
  $\sqrt{\mean{\df_1^2}}+\mean{\df_2}$, $\mean{\df_2}$ being evaluated
  with or without the volume effects.

\par

However, to explicitly derive $\Df_{\max}(\varphi_\cl)$ and
$\Df_{\min}(\varphi_\cl)$, Eq.~(\ref{eqexact}) cannot be used directly
because we do not know the values of $\theta_{\rm L}$ and $\theta_{\rm
R}$. In fact, it is sufficient to take the maximum of the absolute
value of the Lagrange remainders (for $\theta_{\rm L,R} \in [0,1]$) in
order to get an upper bound on the error. Therefore, the approximation
is reliable, \ie $L_2$ and $R_2$ are negligible, when the two
following conditions
\begin{eqnarray}
\label{consl}
\max_{x\in[\varphi_\cl,\,\varphi_\cl+\Df]} \left\vert
  \frac{H^{(4)}(x)}{6}\Df^3\right\vert \ll \left\vert
  \frac{H'''_\cl}{2}\right\vert\Df^2
\end{eqnarray}
and
\begin{equation}
\label{consr}
  \max_{x\in[\varphi _\cl,\,\varphi_\cl+\Df]} \left\vert \frac{
  \left(H^{3/2}\right)''(x)}{2}\right\vert \Df^2 
%\nonumber \\ & & 
  \ll \left\vert
  \left(H^{3/2}_\cl\right)'\Df\right \vert
\end{equation}
hold, while it breaks down when (for fixed values of $\varphi_\cl$) at
least one of the two is violated. The limiting values
$\Df_{\max}(\varphi_\cl)$ and $\Df_{\min}(\varphi_\cl)$ are then
determined by requiring that the two above inequalities become
equalities. Since we have two equations and each of them involves
absolute values, this gives two positive and two negative solutions,
the actual value of $\Df_{\max}(\varphi_\cl)$ and
$\Df_{\min}(\varphi_\cl)$ clearly being the one leading to the
tightest constraint.

\par

Having determined $\Df_{\max}(\varphi_\cl)$ and
$\Df_{\min}(\varphi_\cl)$ with the above procedure, one must also take
into account the fact that we are dealing with stochastic
quantities. In this respect, the validity of the perturbative
treatment will be guaranteed as long as the probability of finding $
-\vert \Df_{\min}(\varphi _\cl)\vert <\df_1+\df_2<\Df_{\max}(\varphi
_\cl)$, is sufficiently close to $1$. In terms of probability, this
means that one requires
\begin{equation}
\label{probaaccuracy}
  \frac{1}{\sqrt{2\pi}\mean{\df_1^2}}
  \int_{\Df_{\min}}^{\Df_{\max}}\kern-1em \dd \varphi\,
  \exp\left[-\frac{\big(\varphi
  -\mean{\df_2}\big)^2}{2\mean{\df_1^2}}\right] \simeq 1 \, ,
\end{equation}
where we have considered $P_{\rm c}$ as the probability density
function. In the case where the volume effects are taken into account,
$P_{\rm v}$ should be used instead.

\par

Finally, there is yet another constraint coming from the fact that, in
general, $\df_1+\df_2$, is a good approximation only if $\df_2\ll
\df_1 $. This is necessary if we want to ``separate''
Eq.~(\ref{eqexact}) into two equations, one for $\df_1$ and one for
$\df_2$.

\par

Let us now see how the previous considerations work in practice for
the chaotic inflation potential $V(\varphi)=m^2\varphi^2/2$. In this
particular case, $\df_1+\df_2$ is an exact solution of the
approximated second order equation since $H'''=0$ and the constraint
$\df_2\ll \df_1$ does not apply. In addition, we also have $L_2=0$
and, as a consequence, the limiting values $\Df_{\min}$ and
$\Df_{\max}$ are found only from the constraint~\eqref{consr}
involving $R_2$. Using the slow-roll equations of motion, one has $
\left(H^{3/2}\right)'(x) =3\left(\kappa /6\right)^{3/4}
m^{3/2}x^{1/2}/2$ and $ \left(H^{3/2}\right)''(x)=3\left(\kappa
/6\right)^{3/4} m^{3/2}x^{-1/2}/4$.
%\begin{eqnarray}
%\left(H^{3/2}\right)'(x) &=& \frac32 \left(\frac{\kappa
%}{6}\right)^{3/4} m^{3/2}x^{1/2} \, ,\\ \left(H^{3/2}\right)''(x) &=&
%\frac34\left(\frac{\kappa }{6}\right)^{3/4} m^{3/2}x^{-1/2}\, .
%\end{eqnarray}
The next step is to evaluate the maximum of this last function in the
interval $x\in [\varphi_\cl,\,\varphi_\cl+\Df]$. Let us start with the
upper bound. Since $\Df_{\max}>0$ one has $\max\left|
\left(H^{3/2}\right)''\right| \propto \varphi _\cl^{-1/2}$, \ie
$\theta _{_{\rm R}}=0$. Then one can solve for $\Df_{\max}$. Applying
Eq.~\eqref{consr}, one arrives at
\begin{eqnarray}
\frac12 &\times & \frac34\left(\frac{\kappa }{6}\right)^{3/4}
m^{3/2}\varphi _\cl ^{-1/2} \Df_{\max}^2 \nonumber \\ &=& \frac32
\left(\frac{\kappa }{6}\right)^{3/4} m^{3/2}\varphi
_\cl^{1/2}\Df_{\max}\, ,
\end{eqnarray}
from which one obtains $\Df_{\max}=4\varphi _\cl$. 

\begin{figure*}[t]
\psfrag{15}{\textsf{1.5}}
  \includegraphics[width=\columnwidth]{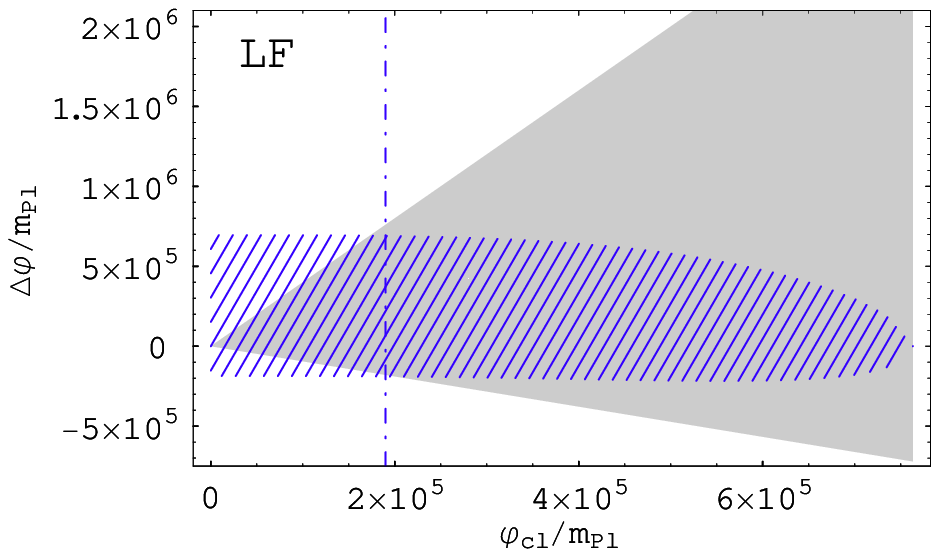}
  \includegraphics[width=\columnwidth]{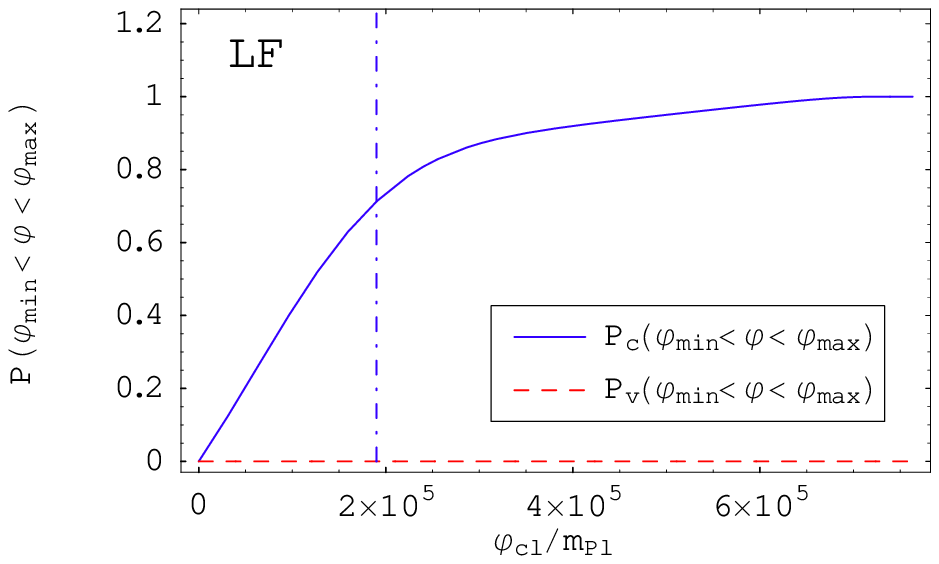}
  \caption{\label{limits_chao} Accuracy of the second order
  approximation for the large field (chaotic) model $V(\varphi)
  \propto m^2\varphi^2$ with an initial condition corresponding to
  $V(\varphi_\ini)=\mpl^4/2$. The mass $m$ is chosen so that the WMAP
  normalization is reproduced (its value is related to the value of
  $M$ used before). All quantities are plotted versus $\varphi_\cl$
  and, therefore, inflation proceeds from the right (large field
  values) to the left (small field values). On the left panel, the
  allowed interval is represented by the uniformly colored (grey)
  region which is delimited by $\Df_{\min}$ and $\Df_{\max}$ obtained
  with Eq.~(\ref{eqrelia}). The hatched (blue) region represents the
  region delimited by the two lines
  $\mean{\df_2}\pm\sqrt{\mean{\df_1^2}}$, $\mean{\df_2}$ being
  evaluated without the volume effects. The vertical dotted-dashed
  (blue) line signals the value of $\varphi _\cl$ at which the
  approximation breaks down. On the right panel, the probability of
  finding $\varphi$ in the reliability range computed with
  $P_\mathrm{c}$ (solid blue line) and with $P_\mathrm{v}$ (red dashed
  line) is displayed. Clearly, the single-domain probability
  probability starts decreasing approximatively at the value of
  $\varphi _\cl$ where $\df_1+\df_2$ is no longer in the reliability
  interval. On the other hand, the volume effects corrections are very
  large and the volume weighted distribution is not trustable.}
  \vspace*{10pt}
  \includegraphics[width=\columnwidth]{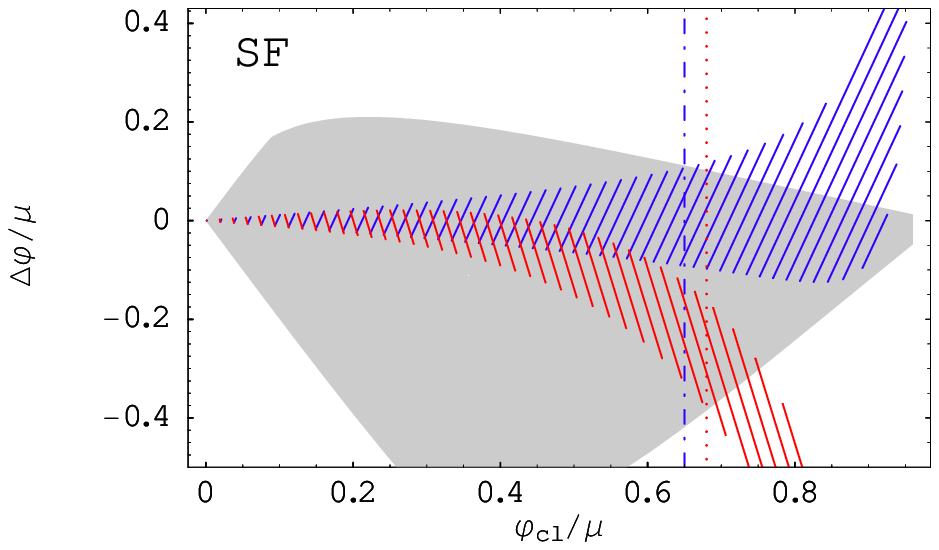}
  \includegraphics[width=\columnwidth]{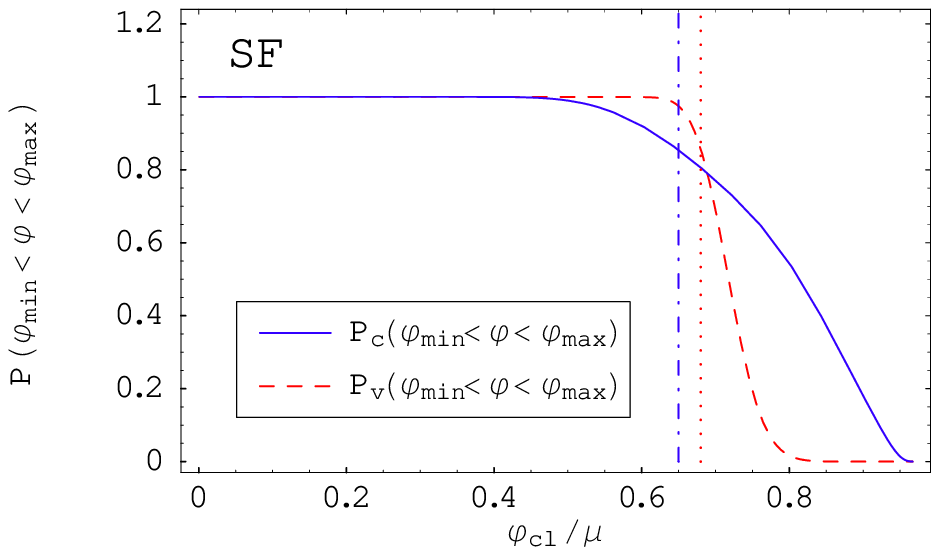}
  \caption{\label{limits_new} Accuracy of the second order
  approximation for the small field (new inflation) potential
  $V(\varphi)\propto 1- (\varphi/\mu)^2$ with the initial condition
  $\varphi_\ini\simeq 10^{-5}\mu$. In this case, inflation proceeds
  from the left (small field values) to the right (``large'' field
  values). On the left panel, the uniformly colored (grey) region
  represents the interval where the approximation is trustable. The
  limiting values are now obtained from the condition on the remainder
  $R_2$ but also from the one coming from the remainder $L_2$. The
  actual $\Df_{\min}(\varphi _\cl)$ and $\Df_{\max}(\varphi _\cl)$,
  which delimit the confidence region, must be the smallest ones (in
  absolute value). The hatched (blue) region with positive slope lines
  is delimited by $\mean{\df_2}\pm\sqrt{\mean{\df_1^2}}$ where
  $\mean{\df_2}$ is computed without the volume effect. On the other
  hand, The hatched (red) region with negative slope lines is also
  delimited by $\mean{\df_2}\pm\sqrt{\mean{\df_1^2}}$ but, this time,
  with $\mean{\df_2}$ computed with the volume effects. The vertical
  dotted-dashed (blue) line indicates when the approximation without
  the volume effects breaks down while vertical dotted (red) line
  signals when the approximation with the volume effects becomes
  untrustworthy. On the right panel, the probability of finding
  $\df_1+\df_2$ in the reliability range is displayed. It is clear
  that both $P_\mathrm{c}$ (solid blue line) and $P_\mathrm{v}$
  (dashed red line) yield a probability close to one for a large part
  of the inflationary phase.}
\end{figure*}

Let us now consider the lower bound $\Df _{\min}<0$. The maximum of
the function $\left(H^{3/2}\right)''$ is now given by $\max\left|
\left(H^{3/2}\right)''\right| \propto \left(\varphi _\cl -\vert
\Df_{\min}\vert \right)^{-1/2}$, \ie $\theta _{_{\rm
R}}=1$. Therefore, in this case, solving the corresponding
equality~\eqref{consr} requires to solve a second order algebraic
equation in $\vert \Df_{\min}\vert $, namely
\begin{eqnarray}
\frac12 &\times & \frac34\left(\frac{\kappa }{6}\right)^{3/4}
m^{3/2}\left(\varphi _\cl -\vert \Df_{\min}\vert \right)^{-1/2} \vert
\Df_{\min}\vert ^2 \nonumber \\ &=& \frac32 \left(\frac{\kappa
}{6}\right)^{3/4} m^{3/2}\varphi _\cl^{1/2}\vert \Df_{\min}\vert\, ,
\end{eqnarray}
and the result reads $\vert \Df_{\min}\vert =(-8\pm 4\sqrt{5})\varphi
_\cl$. Gathering the two limits obtained before, one finds that the
reliability interval is given by
\begin{equation}
\label{eqrelia}
  -4(\sqrt{5}-2)\varphi _\cl \ll \df_1+\df_2 \ll 4\varphi _\cl\, .
\end{equation}
This interval is represented in Fig.~\ref{limits_chao} (left panel) by
the uniformly colored (grey) region. One clearly sees that this region
is limited by two straight lines as calculated above. As inflation
proceeds, the allowed region shrinks. This has to be compared with
$\mean{\df_2}\pm\sqrt{\mean{\df_1^2}}$ represented by the hatched
(blue) area. The lower border line of the hatched (blue) region is
$\mean{\df_2}-\sqrt{\mean{\df_1^2}}$ while the upper border line is
$\mean{\df_2}+\sqrt{\mean{\df_1^2}}$. The two lines meet at the
beginning of inflation where they vanish since $\mean{\df_1^2(\varphi
_\ini)} =\mean{\df_2(\varphi _\ini)}=0$. As long as the hatched region
lies within the uniformly colored one, the approximation is
reliable. When this is no longer the case, the approximation breaks
down. In Fig.~\ref{limits_chao}, this is signaled by the vertical
dotted-dashed line and occurs for $\varphi _\cl \sim 2\times
10^{5}\mpl$. It is clear that the second order approximation is good
until one approaches the end of slow-roll inflation. The right panel
of the same figure shows the probability of finding $\df_1+\df_2$
between $\Df_{\min}$ and $\Df_{\max}$, computed according to
Eqs.~(\ref{probaaccuracy}) and ~(\ref{eqrelia}), and confirms the
previous conclusion.

\par

When volume effects are considered, the situation becomes conceptually
more complicated but the same ideas can be utilized to check the
accuracy of the volume-weighted distribution. In particular, one
should now compare the region limited by $\mean{\df_2}+3I^{\rm
T}J\pm\sqrt{\mean{\df_1^2}}$ with the reliability region. In the case
of large field models, however, we do not plot this region because the
volume effects are so important that the corresponding region would be
outside the figure. This will be done for the other models, see below.

\par

One can also use the criterion of Eq.~(\ref{probaaccuracy}) but this
time, as already mentioned above, the probability should be evaluated
with the distribution $P_\mathrm{v}(\varphi)$ rather than with
$P_\mathrm{c}(\varphi)$. The corresponding results can strongly differ
since the field realizations having higher potential energy (\ie with
a faster expansion rate) will be favored. In particular, if their
expansion rate is sufficiently large, this can give a high statistical
significance to realizations outside the reliability range having a
very low significance according to the original distribution. In this
situation, when the difference between the two distributions is very
important, the form of $P_\mathrm{v}(\varphi)$ obtained from the
perturbed solution cannot be trusted although $P_\mathrm{c}(\varphi)$
is reliable. This means that, in most domains, the statistical
properties of the field are correctly described by
$P_\mathrm{c}(\varphi)$ but that the Universe is mainly made up of
very big domains where $P_\mathrm{c}(\varphi)$ cannot be trusted. This
is exactly what happens for LF models where the volume-weighted
probability of finding the field in the confidence range (dashed red
line, right panel in Fig.~\ref{limits_chao}) is basically vanishing
while the single-domain one (solid blue line) is large. Therefore, in
this case, the perturbative method does not allow us to reliably
compute the volume-weighted distribution.

\begin{figure*}
  \includegraphics[width=\columnwidth]{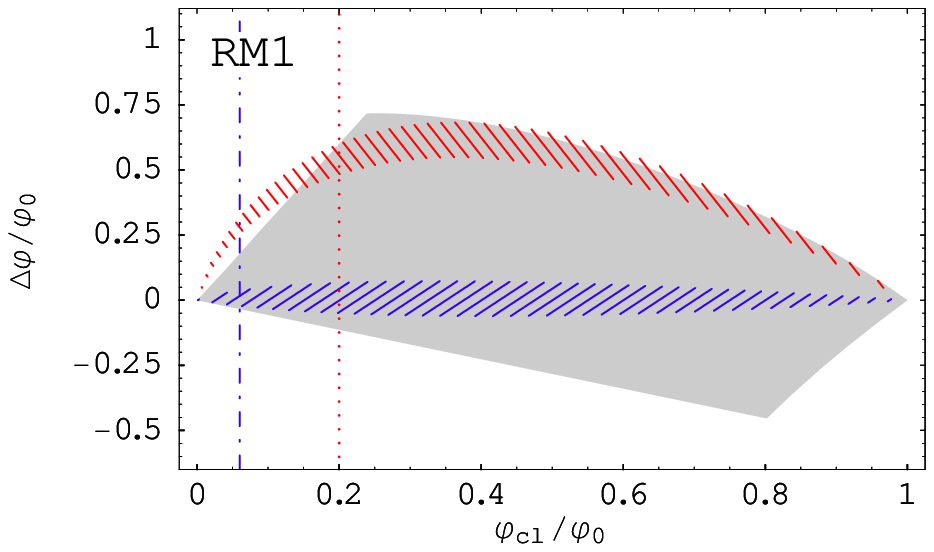}
  \includegraphics[width=\columnwidth]{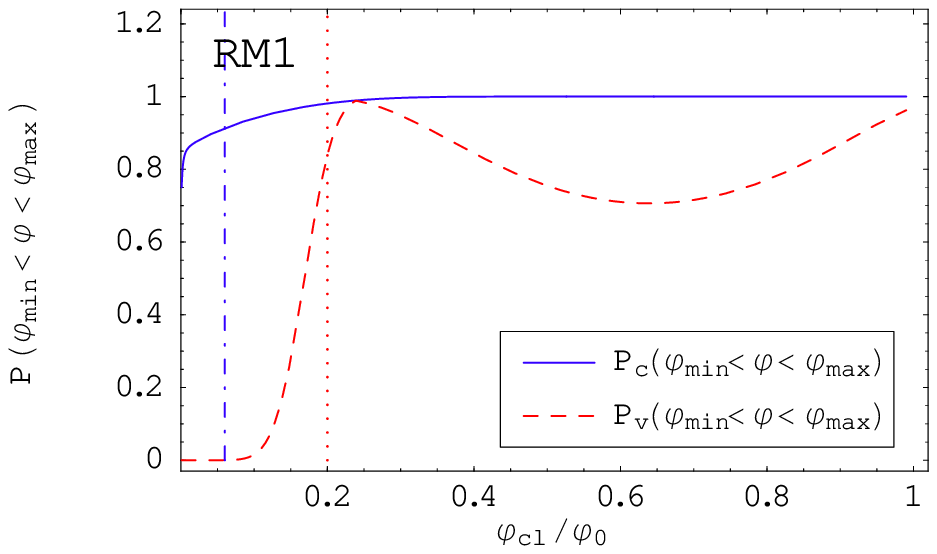}
\\
  \includegraphics[width=\columnwidth]{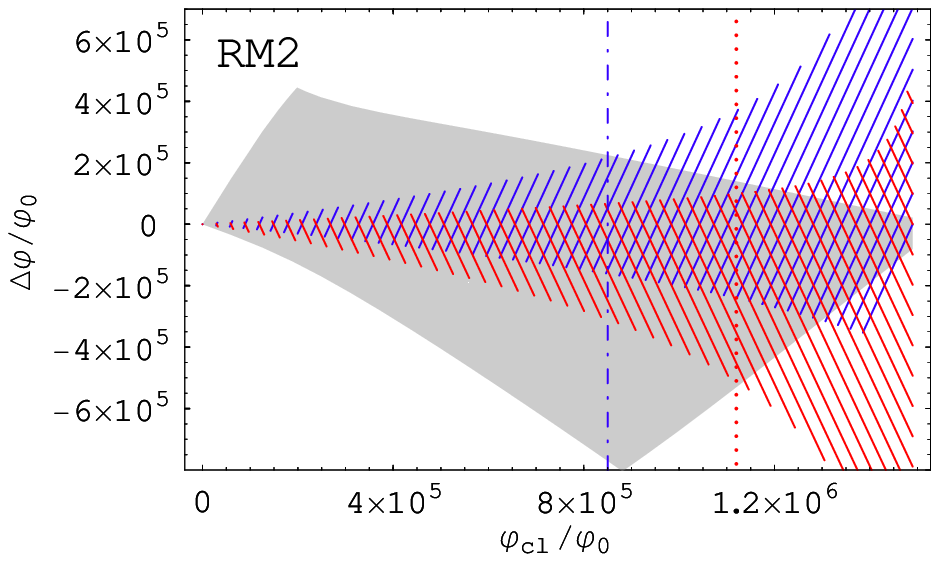}
  \includegraphics[width=\columnwidth]{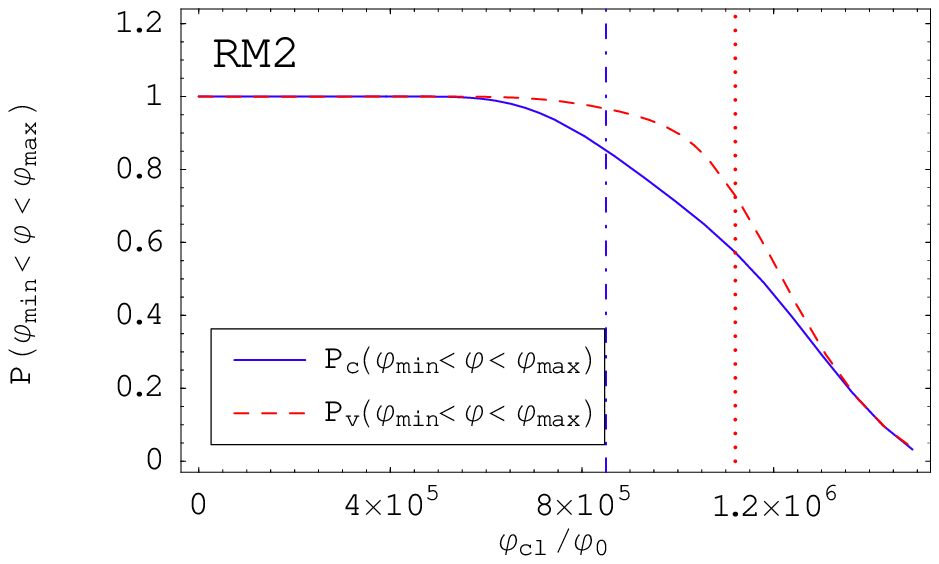}
  \caption{\label{limits_run} Accuracy of the second order
  approximation for the two running mass models RM1 and RM2 with
  initial conditions $\varphi_\ini/\varphi _0\simeq 1-1.5\times
  10^{-5}$ and $\varphi_\ini/\varphi _0\simeq 1+10^{-3}$
  respectively. Inflation proceeds from the right to the left in the
  RM1 model (upper panels) and from the left to the right in the RM2
  model (lower panels). On the left panels, the limiting values
  signaling the break down of the conditions given by
  Eqs.~\eqref{leftcond} and~\eqref{rightcond} are represented and
  compared with $\mean{\df_2}\pm\sqrt{\mean{\df_1^2}}$. Conventions
  are the same as the ones used in Figs.~\ref{limits_chao} and
  \ref{limits_new}. As usual, the actual values of $\Df_{\min}$ and
  $\Df_{\max}$ must be the ones with the smallest absolute
  values. Cusps in the curves are a consequence of taking the maximum
  absolute value of the Lagrange remainders and appear when one
  discontinuously change the value of $\theta $ from $0$ to $1$ (or
  the opposite). On the right panels, the probability of finding
  $\varphi$ in the reliability range is displayed. For both models the
  reliability of the solution does not dramatically change (and for
  the RM2 model is even slightly enhanced) when volume weighting is
  considered}
\end{figure*}

We have also performed the same study for new inflation and the
results are displayed in Fig.~\ref{limits_new}. Now, the reliability
interval, still given by the uniformly colored (grey) region, is no
longer determined only from $R_2$ but also from $L_2$ because,
contrary to the LF case, $L_2$ does not vanish for SF. Therefore, in
this situation, one has to determine $\Df_{\max}$ and $\Df_{\min}$
from Eqs.~(\ref{consl}) and (\ref{consr}) and not only from
Eq.~(\ref{consr}), as was the case for the LF models. The form of
$\Df_{\max}$ and $\Df_{\min}$ as a function of $\varphi _\cl $ is also
more complicated and is no longer given by straight lines. It must be
computed numerically. As can be seen in Fig.~\ref{limits_new}, the
allowed region increases at the beginning of inflation, reaches a
maximum extension and shrinks as the end of inflation is
approached. The region $\mean{\df_2}\pm\sqrt{\mean{\df_1^2}}$ without
the volume effects is given by the hatched (blue) region, the lines
having a positive slope.  The region
$\mean{\df_2}\pm\sqrt{\mean{\df_1^2}}$ with the volume effects taken
into account is represented by the hatched (red) region, the lines
having a negative slope. When these regions are within the uniformly
colored (grey) region, the approximation is reliable.  As before, the
border lines $\mean{\df_2}\pm \sqrt{\mean{\df_1^2}}$ meet at the
beginning of inflation where they vanish. The break down of the
approximation, without the volume effect, is signaled by the vertical
dotted-dashed (blue) line and, with the volume effects, by the dotted
(red) line. One notices that the results for SF are basically similar
to those that have been obtained in the chaotic model case, namely the
approximation remains reliable until the very end of the inflationary
phase. However, in the case of new inflation, one clearly sees the
importance of using a carefully defined criterion to estimate the
reliability of the approximation. As already mentioned, the naive
criterion $\df_1 \ll \varphi _\cl $ would have indicated that the
approximation becomes untrustworthy very quickly after the beginning
of inflation since we have initially $\varphi _\cl/\mu \ll 1$. We see
in Fig.~\ref{limits_new} that, on the contrary, the approximation is
good during a large part of inflation. Finally, the probability of
being in the reliability region, computed with $P_{\rm c}$ (solid blue
line) or $P_{\rm v}$ (dashed red line) is also displayed in
Fig.~\ref{limits_new} (right panel). In the case of SF models, the
volume effects are less important and, as a consequence, the two
probabilities are similar. The fact that the break down of the
approximation occurs at the end of inflation only, at $\varphi
_\cl/\mu \sim 0.65$ for the particular example studied here, is
confirmed.

\par

A similar analysis can also be done for the running mass
potential. The results for the two models under consideration are
displayed in Fig.~\ref{limits_run}.  All the conventions concerning
the allowed regions, volume effects etc \dots are the same as
before. In the case of the RM1 model (upper panels), Eq.~\eqref{consl}
giving the constraint on $L_2$ can be analytically solved to the
lowest order in $c(\varphi_0/\MPl)^2$. This gives
\begin{equation}
  -\frac{7-\sqrt{13}}{6}\varphi_\cl \ll \Df \ll 3\varphi_\cl \, .
\end{equation}
It turns out that this constraint is the dominating one at late times
as can be checked in Fig.~\ref{limits_run} (left panel): the shape of
the allowed region near the end of inflation is delimited by two
straight lines. The constraint on $R_2$ has been solved
numerically. The figure demonstrates, and this is also confirmed in
the right panel, that the perturbative solution for RM1 is very good
during almost all the inflationary phase and breaks down only at the
very end of inflation. Similar conclusions hold for the RM2 model, see
the two bottom panels.

\par

Finally, one notices that the two probabilities [\ie the ones obtained
with $P_\mathrm{c}(\varphi)$ and $P_\mathrm{v}(\varphi)$], and
contrary to the LF case, do not dramatically differ from each
other. In the case of the RM2 model, the reliability of the
volume-weighted description can even be larger than the single-point
one.

\section{Conclusions}
\label{Conclusions}

In this section, we briefly summarize the new results obtained in this
article. The main goal of the paper was to present a new method aimed
at estimating the precision of the perturbative expansion studied in
Ref.~\cite{MM2}. This method is based on the use of the Lagrange
remainder. After having discussed the general features of this new
approach, we have applied it to the inflationary models studied in
Ref.~\cite{MM2}. We have proven that the approximate probability
density functions derived in this reference are, in general, a very
good approximations to the actual ones except, as expected, at the end
of inflation. This conclusion holds even if the volume effects are
taken into account except in the case of the large field models. We
conclude that the perturbative expansion of the inflationary Langevin
equation together with the method presented here, besides being the
only available method with a built-in measure of its domain of
validity, form a robust formalism to efficiently compute the
stochastic effects during inflation.

\section{Acknowledgments}

M.~M. would like to thank the Universit\`a degli Studi di Milano and
the Institut d'Astrophysique de Paris (IAP) where part of this work
has been done. We would also like to thank Sergio Cacciatori for
enlightening discussions and Patrick Peter for careful reading of the
manuscript.

The work of M.~M. is supported by the National Science Foundation 
under Grant Nos. PHY-0071512 and PHY-0455649, and with grant support 
from the US Navy, Office of Naval Research, Grant Nos. N00014-03-1-0639 
and N00014-04-1-0336, Quantum Optics Initiative.

\end{document}